\documentclass[12pt]{article}

\usepackage{amsmath,amssymb}
\usepackage{graphicx}
\usepackage{xspace}
\usepackage[sf,SF]{subfigure}
\usepackage[format=plain,labelfont=bf,justification=RaggedRight,font=small]{caption}

\usepackage{amssymb}
\usepackage{comment}
\usepackage[utf8]{inputenc}

\usepackage{subfigure}
\usepackage{color}

\newcommand{\be}{\begin{equation}}
\newcommand{\ee}{\end{equation}}

\definecolor{greenvy}{RGB}{0,128,0}
\definecolor{bluevy}{RGB}{0,191,191}
\definecolor{violetvy}{RGB}{191,0,191}
\definecolor{orangevy}{RGB}{255,150,0}

\title{Contact between representative rough surfaces}

\author{\small Vladislav A. Yastrebov$^{1,2}$\footnote{vladislav.yastrebov@mines-paristech.fr}, Guillaume Anciaux$^2$\footnote{guillaume.anciaux@epfl.ch}, Jean-Fran{\c c}ois Molinari$^2$\footnote{jean-francois.molinari@epfl.ch}\\[5pt]
\footnotesize$^1$\footnotesize Centre des Mat\'eriaux, MINES ParisTech, CNRS UMR 7633,\\\footnotesize BP 87, 91003, Evry, France\\[3pt]
\footnotesize$^2$\footnotesize Computational Solid Mechanics Laboratory (LSMS, IIC-ENAC, IMX-STI),\\\footnotesize Ecole Polytechnique F{\'e}d{\'e}rale de Lausanne (EPFL),\\ \footnotesize Station 18, 1015, Lausanne, Switzerland}
\date{\footnotesize\today}

\begin{document}

\maketitle
\hrule\vspace{0.3cm}

\begin{flushleft}\large\textbf{Abstract}\end{flushleft}\normalsize
A numerical analysis of mechanical frictionless contact between rough self-affine elastic manifolds was carried out. It is shown that the lower cutoff wavenumber in surface spectra is a key parameter controlling the representativity of the numerical model. Using this notion we demonstrate that for representative surfaces the evolution of the real contact area with load is universal and independent of the Hurst roughness exponent. By introducing a universal law containing three constants,
we extend the study of this evolution beyond the limit of infinitesimal area fractions.

\vspace{0.3cm}\hrule\vspace{0.3cm}
\section{Introduction}

Real surfaces are self-affine \cite{meakin1998b,krim1995ijmpb} and the surface heights are distributed normally~\cite{greenwood1966prcl}.
Due to this roughness the real contact area $A$ is often only a small fraction of the nominal or apparent contact area $A_0$.
The real contact area fraction $A/A_0$ and its evolution determines the contact resistivity, friction and the
transfer of energy (heat and electric charge) through the contact interface. Thus understanding the contact-area evolution has profound implications
in various fundamental (e.g., origin of friction) and engineering studies (e.g., electro-mechanical contact, tire-road interaction).
From early experiments~\cite{bowden2001b} and analytical theories \cite{greenwood1966prcl,bush1975w}, it has been considered as an established knowledge that 
$A$ evolves linearly with the normal load $F$ at relatively small fractions of contact.
It was confirmed by numerical simulations of normal frictionless elastic contact between rough surfaces~\cite{hyun2004pre}, which made considerable progress since then.
Nowadays, the most advanced computations exploit fine surface discretizations ($2048\times2048$ \cite{campana2007epl,campana2011jpcm,almqvist2011jmps,pohrt2012prl}, 
$4096\times4096$~\cite{campana2008jpcm})
to extract statistically meaningful results valid even for small fractions of contact. 
Many of the numerical investigations are based on synthesized surfaces preserving the aspect of self-affinity.
In early studies \cite{hyun2004pre,pei2005jmps} the artificial rough surfaces obeyed fractality which scaled down to the domain discretization \cite{fournier1982cacm}. 
Later it was realized that the surface has to be smooth enough \cite{campana2007epl,yastrebov2011cras} 
to represent correctly the mechanics of contact and to obtain a reliable estimation of the contact area growth comparable with analytical theories.
This smoothness implies a sharp decay of the power spectrum density for wavelengths smaller than 
short cutoff wavelength $\lambda_s$.
On the other hand, to the best of our knowledge, there are no consistent study on the influence of long cutoff wavelength $\lambda_l$ on the response of rough surfaces. 
Despite remarks made in~\cite{hyun2007ti} on the importance of $\lambda_l$ for mechanics of rough contact, 
in almost all investigations of the real-contact-area evolution \cite{hyun2004pre,campana2007epl,pohrt2012prl,putignano2012jmps} 
$\lambda_l$ was limited by the size of the specimen $\lambda_l=L$.
In this article we show that $\lambda_l$ plays a crucial role in numerically precise theory of rough contact, because 
it controls the representativity of the simulated system.

\begin{figure}
   \begin{center}
   \includegraphics[width=0.8\textwidth]{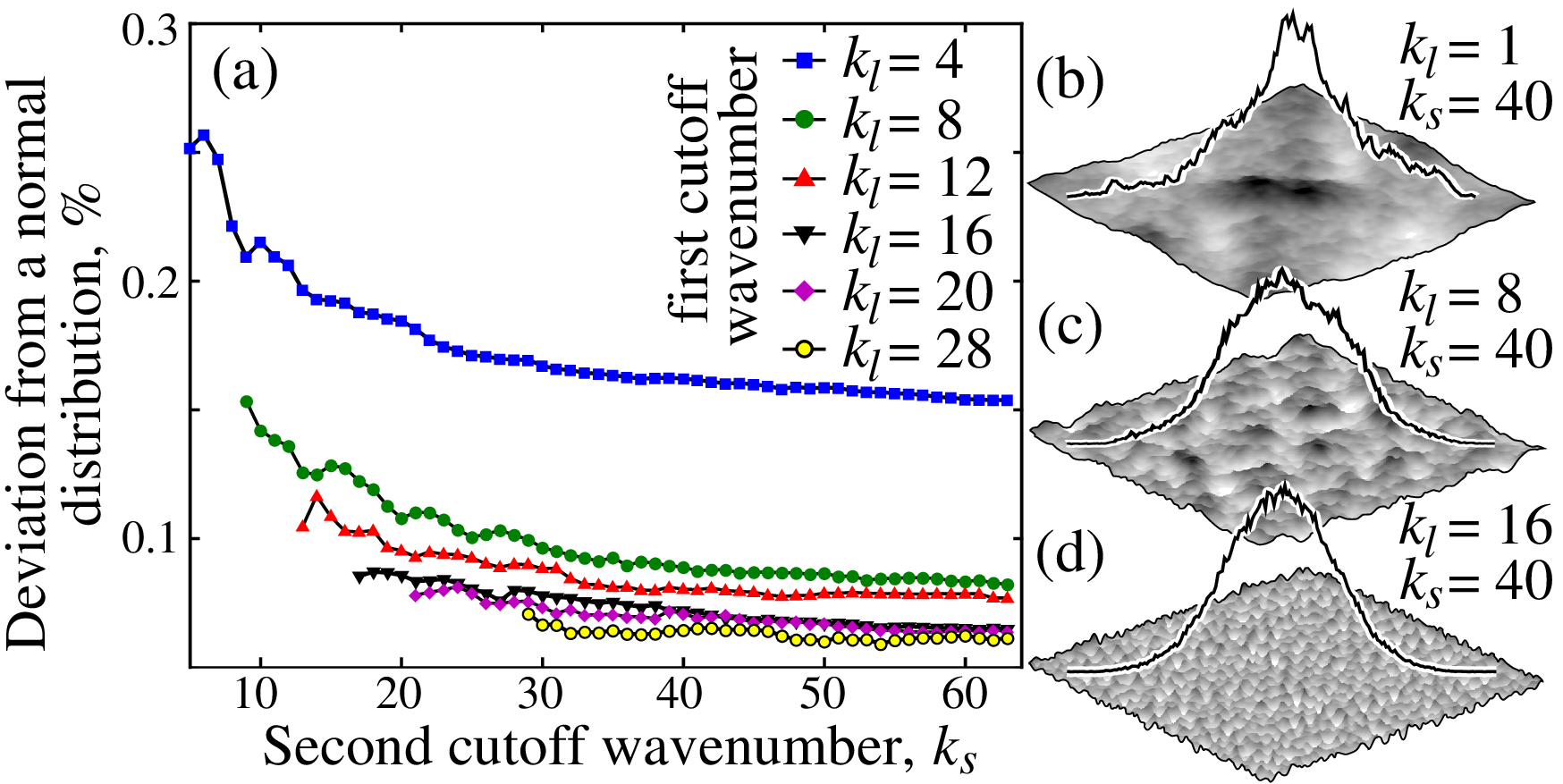}
   \caption{\label{fig:1}
Representativity of a rough self-affine surface can be estimated according to the normality of heights distribution.
Here, the criterion of proximity is an error between the distribution of surface's heights and the associated
normal distribution; 
(a) this error is strongly dependent on the cutoff wavenumbers $k_s$ and $k_l$.
Every point on the graph represents an error averaged over 30 statistically equivalent surfaces.
Examples of rough surfaces and the corresponding heights distributions for different cutoffs $k_s=40$ and (b) $k_l=1$, (c) $k_l=4$, (d) $k_l=16$ are also depicted.}
   \end{center}
\end{figure}
In the limit of infinitesimal contact, analytical theories predict that the real contact area $A$ evolves proportionally to the normal force $F$ with coefficient $\kappa$ normalized by 
the root mean squared slope of the surface $\sqrt{\langle |\nabla h|^2\rangle}$ and the effective Young's modulus $E^*$~\cite{johnson1987b}, which gives  
\begin{equation}
A = \kappa/\sqrt{\langle |\nabla h|^2\rangle} \;F/E^*.
\label{eq:0}
\end{equation}
According to asperity-based~\cite{greenwood1966prcl} models (Bush, Gibson, Thomas (BGT)~\cite{bush1975w}, Greenwood~\cite{greenwood2006w}, Nayak-Thomas~\cite{thomas1982b})
$\kappa = \sqrt{2\pi}$ but Eq.~(\ref{eq:0}) is valid only in an asymptotic limit for infinitesimal fractions of the contact area $A/A_0\to0$.
The convergence of $\kappa$ with decreasing fraction $A/A_0$ is slow and depends
on the bandwidth parameter $\alpha$ introduced in~\cite{longuet-higgins1957rsla} as $\alpha = m_0 m_4/m_2^2$, where $m_i = \int_{k_l}^{k_s}k^iC(k) dk$ is the $i$-th moment of the power spectrum density $C(k)$ of the surface.
For instance, for $\alpha=10$, $\kappa$ is overestimated by about $12$\% for $A/A_0=10^{-5}$ 
and this error reduces only to $4$\% when $A/A_0=10^{-10}$~\cite{greenwood2006w}.
Consequently, using a complete numerical model it is not possible to demonstrate the linearity predicted by Eq.~(\ref{eq:0}) nor to approach its asymptotic limit.
It is also worth noting that in forementioned models the evolution of the real contact area is strictly nonlinear for realistic fractions of contact~\cite{carbone2008jmps,paggi2010w}. 
Moreover, this evolution does not depend on the Hurst roughness exponent $H$ but only on $\alpha$~\cite{carbone2008jmps}.
A competing theory was proposed by Persson~\cite{persson2001jcp,persson2001prl}, it also predicts a linear contact area evolution for small 
contact fractions, however, the obtained coefficient of proportionality $\kappa = \sqrt{8/\pi}$ is significantly smaller than in the asperity-based models. 
As in the latter models, the evolution of the contact area does not depend on the Hurst exponent.
Numerical studies \cite{hyun2004pre,campana2007epl,pohrt2012prl,putignano2012jmps}
demonstrated an approximately linear evolution of the contact area with load. 
But in contrast to the analytical models, $\kappa$ was shown to depend on the Hurst exponent and to be confined 
between the asymptotic limits of the BGT~\cite{bush1975w} and Persson~\cite{persson2001jcp,persson2001prl} theories.
In this article, by introducing the notion of surface \emph{representativity}, we obtain qualitatively new numerical results, which we believe
correctly represent the mechanical response of realistic rough surfaces.
Moreover, they appeal to a broad interpretation of the data not restricted to a single proportionality coefficient between contact area and load for infinitesimal
contact fractions.
\begin{figure}
   \begin{center}
   \includegraphics[width=0.8\textwidth]{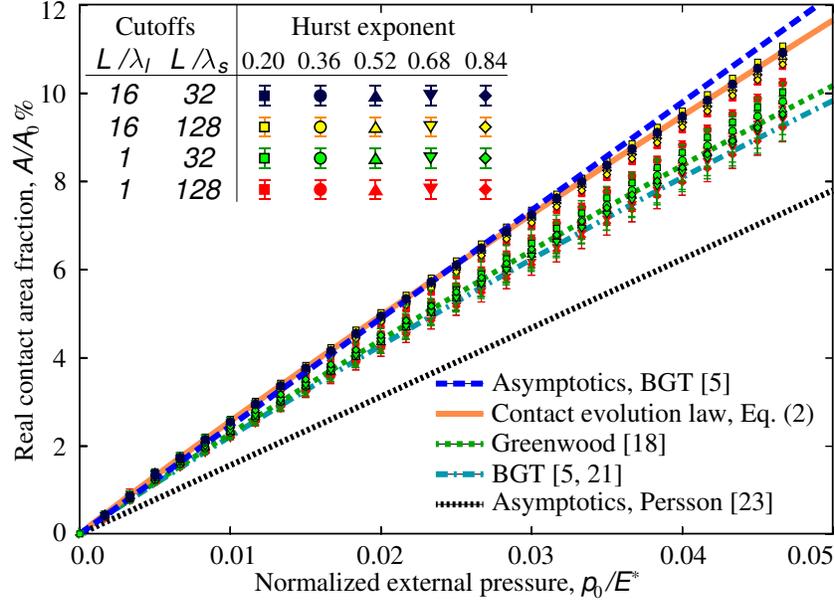}
   \caption{\label{fig:5}
Evolution of the real contact area fraction $A/A_0$ with normalized external load $p_0/E^*$
are plotted for different cutoff wavelengths and Hurst exponents; the rms slope is kept constant for all surfaces $\sqrt{\langle |\nabla h|^2\rangle} = 0.1$. 
The results for representative surfaces ($\lambda_l=L/16$) are in a good agreement with the asymptotic limit of the Bush, Gibson, Thomas (BGT) theory~\cite{bush1975w} and are perfectly fitted by the suggested contact evolution law Eq.~(\ref{eq:1}) ($\kappa=1.145\sqrt{2\pi},\mu=0.55,\beta=0.21$). 
Moreover, if the surface is sufficiently smooth $\lambda_s = L/32$ we do not observe any dependence of results on the Hurst exponent. 
The evolution of the real contact area predicted by Greenwood~\cite{greenwood2006w} and BGT~\cite{bush1975w} theories for spectrum bandwidth $m_0 m_4/m_2^2=2$ 
(data from~\cite{carbone2008jmps}) lie in the confidence interval of the results obtained for non-representative surfaces ($\lambda_l=L$) 
with high Hurst exponents $H\gtrapprox0.5$.
Asymptotic limit of Persson theory~\cite{persson2001jcp,persson2001prl} lies below all obtained results.
}
   \end{center}
\end{figure}
\section{Representativity of rough surfaces}

A meaningful numerical simulation of rough contact has to be carried out on 
a representative self-affine surface element (RSSE) (e.g.~\cite{kanit2003ijss})
either generated numerically or chosen from experimental measurements.
The reason to use generated rough surfaces is that they 
may be obtained for any cutoff wavelengths and Hurst exponent $H$.
On the one hand, the RSSE has to be large enough to obtain a similar response for different realizations of statistically equivalent surfaces. 
On the other hand, it has to be as small as possible to retain a numerically solvable contact problem.
So the RSSE is defined according to the permitted error between mechanical responses of surfaces of different sizes, e.g. $L$ and $2L$.
This mechanical representativity can be linked to the geometrical representativity, 
which we define as the proximity of 
the surface heights distribution to a normal distribution~\cite{yastrebov2011cras}. 
In particular, this proximity is important in the range of maximal heights corresponding to the only zones which come in contact at small loads.
To analyze how the representativity depends on the cutoffs, for each pair of cutoff wavenumbers $k_s=L/\lambda_s$ and $k_l=L/\lambda_l$ we generate 30 statistically equivalent surfaces
using FFT filtering algorithm~\cite{hu1992ijmtm}. The size of surfaces is $L=1$, the discretization spacing is $\Delta L = 1/1024$ and $H=0.8$.  
Fig.~\ref{fig:1} represents the average $L2$ error between the heights distribution function of the generated surfaces and the corresponding normal distributions $G(h,\bar h,\sigma)$ for different cutoffs (see appendix for details).
The mean value $\bar h$ and the standard deviation $\sigma=\sqrt{\langle(\bar h - h)^2\rangle}$ for the normal distribution $G$ 
are computed according to the generated surface.
It follows from this figure that a generated surface becomes more and more representative, when the first cutoff wavenumber $k_l$ increases up to $8-16$,
whereas the variation of $k_s$ has a weaker effect on the surface representativity if the spectrum is sufficiently rich $k_s \gg k_l$.
The choice $\lambda_l=L$ leads to a surface with a mechanical response which varies considerably from one surface realization to another.
The fact that in average this response does not correspond to the response of a bigger rough surface with similar random properties is of
crucial importance for the study of rough contact.
The key reasons for this discrepancy are the long-range interactions ($1/r$) between contacting asperities and 
inevitable boundary conditions on lateral sides of the specimen (periodic, symmetric, free).
Accordingly, the response of any heterogeneous system with random properties and long-range interactions should be 
studied on a representative system element.
\section{Smoothness of rough surfaces}
\begin{figure}
   \begin{center}
   \includegraphics[width=0.8\textwidth]{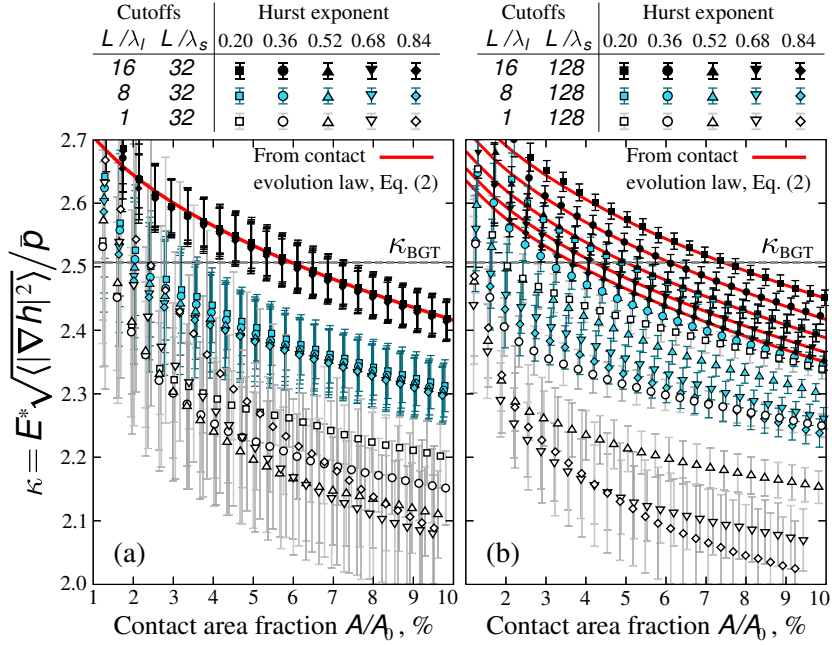}
   \caption{ \label{fig:3}
Evolution of $\kappa$ with the real contact area fraction $A/A_0$ is depicted for different Hurst roughness exponents and cutoff wavelengths $\lambda_l=L,\,L/8,\,L/16$, a) $\lambda_s=L/32$, b) $\lambda_s=L/128$; 
all results for representative surfaces $\lambda_l=L/16$ are fitted by the contact evolution law Eq.~(\ref{eq:1}) with $\kappa=1.145\sqrt{2\pi},\;\mu=0.55,\;\beta=0.21$ for
(a) $\lambda_l=L/16$, $\lambda_s=L/32$; with $\kappa=1.145\sqrt{2\pi},\;\mu=0.56$ and $\beta \in [1.120,\;1.172]$ for (b) $\lambda_l=L/16$, $\lambda_s=L/128$.
For comparison, the asymptotic limit $\kappa_{\mbox{\tiny BGT}}=\sqrt{2 \pi}$ from Bush, Gibson, Thomas theory~\cite{bush1975w} is plotted, 
while the asymptotic limit of Persson theory~\cite{persson2001jcp,persson2001prl} $\kappa_{\mbox{\tiny P}}=\sqrt{8/\pi}\approx1.60$ is out of the plot range.}
   \end{center}
\end{figure}

Besides the long cutoff wavelength, in generated or experimental surfaces
there is a limitation connected with their inevitable discreteness, either 
due to a numerical resolution scheme and/or experimental measurements.
In real self-affine surfaces the power spectrum density decays as a power-law of the wavenumber $k$, $C(k) \sim k^{-2(1+H)}$. 
This law may be preserved down to wavelengths comparable to atomic spacings~\cite{krim1995ijmpb}, 
where the continuum contact mechanics is not valid anymore~\cite{robbins2005n}.
To remain in the continuum framework and to capture accurately the mechanics,
we introduce a short cutoff wavelength $\lambda_s > \Delta L$, which ensures the smoothness of surfaces at a certain magnification as, for example, in~\cite{campana2007epl}.
The inherent cutoff $\lambda_s = \Delta L$ used, for example, in \cite{hyun2004pre,pei2005jmps} results in a ``one node - one asperity'' approach, which does not allow
to correctly reproduce the local change of the contact force with separation~\cite{yastrebov2011cras} nor 
to estimate a realistic growth of localized contact zones but only the growth of their number.
Considering smooth--rough-surfaces with a ``truncated self-affinity'' $\lambda_s \gg \Delta L$ eliminates these shortcomings. 
\section{Description of simulations}

Using the algorithm~\cite{hu1992ijmtm} we generate 12 statistically equivalent surfaces for each pair of cutoff wavelengths 
$L/\lambda_l = 1, 2, 8, 16$, $L/\lambda_s = 32, 64, 128, 256, 512$ and different Hurst roughness exponents $H=0.2, 0.36, 0.52, 0.68, 0.84$. 
For all the surfaces the constant root mean squared slope $\sqrt{\langle |\nabla h|^2\rangle} = 0.1$ is preserved; 
as before $L=1$ and $\Delta L = 1/1024$.
It is important to note that, in contrast to geometrical estimations of $\sqrt{\langle |\nabla h|^2\rangle}$~\cite{hyun2004pre},
which depend on the surface discretization and thus may underestimate significantly the real value, we evaluate the rms slope according to the surface power spectrum~\cite{longuet-higgins1957rsla}.
To solve the contact problem between a periodic rigid rough surface and a deformable flat half-space we use the spectral based boundary element method~\cite{stanley1997} with some minor improvements~\cite{commentKato}.
This method allows to solve accurately the equations of continuum mechanics under contact constraints and, in contrast to asperity-based models, takes into account all underlying mechanics: complex shapes of asperities, junction of contact zones associated with different asperities and
long-range deformation of the elastic half-space in response to contact forces.
For each simulation the contact area fraction reaches $\approx 10\%$ under the external pressure linearly increasing up to $p_0/E^*=0.05$ within $30$ increments.
In Fig.~\ref{fig:5} we compare the area-force curves of our numerical results and several analytical theories,
which were computed and summarized in~\cite{carbone2008jmps}.
To demonstrate better the nonlinearity of the area evolution for surfaces with different cutoffs and Hurst exponents,
we present in Fig.~\ref{fig:3} the proportionality coefficient $\kappa$ expressed from Eq.~\ref{eq:0} 
as $\kappa = E^* \sqrt{\langle |\nabla h|^2\rangle}/\bar p$, where $\bar p = F/A$ is the mean contact pressure.
\section{Results}

The value of $\kappa$ for the observed range (up to 10\% of the contact fraction) is a decreasing function of the area.
For non-representative $\lambda_l < L/16$ or too rough $\lambda_s < 32\Delta L$ surfaces,
there is a clear tendency (see Fig.~\ref{fig:3}, b): a higher $H$ results in a smaller $\kappa$.
This dependence of the results on the Hurst exponent corresponds to all up to date numerical investigations of the $\kappa$ constant \cite{hyun2004pre,campana2007epl,putignano2012jmps,pohrt2012prl}.
However, we argue that these results are strongly affected by non-representativity and an excessive roughness of exploited surfaces.
In contrast to these results, if the mechanics of contact is well resolved ($\lambda_s \gg \Delta L$) and the surface
is representative ($\lambda_l \gtrapprox L/16$) 
the real contact area evolution does not depend anymore on the Hurst exponent (see dark symbols in Fig.~\ref{fig:3}, a).
This result is in a good agreement with analytical theories and has never been obtained before in complete numerical models.
Reducing $\lambda_l$ reduces the data scatter, which is one of criteria of mechanical representativity.
The data scatter increases with increasing $H$, thus, the smoother the surface, the bigger RSSE is needed to capture an average mechanical behavior.
The rise in $\kappa$ with decreasing $\lambda_l$ 
can be interpreted in terms of Persson theory~\cite{persson2001jcp,persson2001prl}:
reducing $\lambda_l$ is equivalent to decreasing magnification which rises $\kappa$.
Note that our results for $\kappa$ are not confined between the asymptotic limits of 
the BGT~\cite{bush1975w} $\kappa_{\mbox{\tiny BGT}}=\sqrt{2\pi}$ and Persson theories $\kappa_{\mbox{\tiny P}}=\sqrt{8/\pi}$~\cite{persson2001jcp}, 
as was generally observed by other authors \cite{hyun2004pre,campana2007epl,putignano2012jmps,pohrt2012prl}.

\section{Phenomenological Contact Evolution Law}

\begin{figure}
   \begin{center}
   \includegraphics[width=0.8\textwidth]{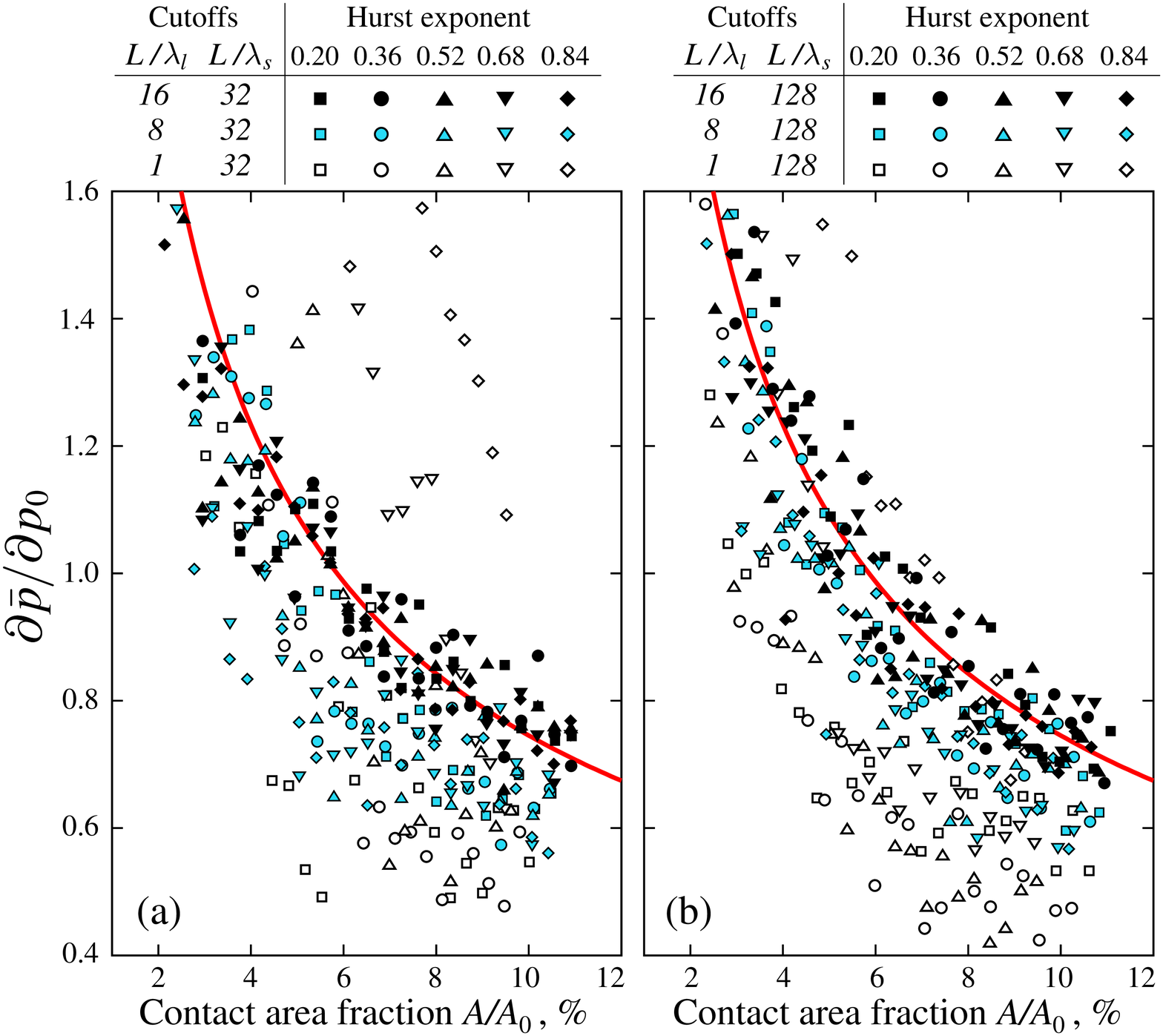}
   \caption{\label{fig:4}
The mean pressure derivative with respect to the external pressure $\partial \bar p/\partial p_0$ 
decreases nonlinearly with increasing fraction of the real contact area $A/A_0$.
We plot its evolution for different Hurst roughness exponents 
and cutoff wavelengths $\lambda_l=L,\,L/8,\,L/16$ and (a) $\lambda_s=L/32$, (b) $\lambda_s=L/128$. 
For representative surfaces the results can be roughly approximated by a power law
$\partial \bar p/\partial p_0 = \beta (A_0/A)^\mu$; solid line
corresponds to parameters $\beta=0.21, \mu=0.55$, which fit well our results for
representative $\lambda_l=L/16$ smooth surfaces.
}
   \end{center}
\end{figure}

In light of these results, we propose a new phenomenological contact evolution law which is
based on our observations of the change in the mean contact pressure $\bar p = F/A$ with respect to the applied pressure $p_0 = F/A_0$ depicted in Fig.~\ref{fig:4}.
For considered interval of contact areas $A/A_0 \in [0.01,0.11]$ the decay of $\partial \bar p/\partial p_0$ can be approximated by power-law 
$\partial \bar p/\partial p_0 = \beta (A_0/A)^\mu$, $0 < \mu < 1$.
Therefore, we obtain (for details see appendix the evolution of the real contact area as
\begin{equation}
  \label{eq:1}
  A/A_0 = \left(\beta +  \left[\kappa p_0/(\sqrt{\langle|\nabla h|^2\rangle} E^*)\right]^{\mu-1}\right)^{1/(\mu-1)}
\end{equation}
where $\kappa$ is still a proportionality constant between the area and normalized force for infinitesimal pressures,
two other constants $\mu$ and $\beta$ can be easily and uniquely found from the data analysis.
For converged results $\lambda_l=L/16$, $\lambda_s=L/32$ by fitting the normalized inverse mean pressure (Fig.~\ref{fig:3}) we found $\kappa = (1.145\pm0.002) \sqrt{2\pi}$,
which is about 15\% higher than the asymptotic limit of the BGT theory~\cite{bush1975w}. 
From Fig.~\ref{fig:4} we uniquely obtain $\mu=0.55\pm0.02$, $\beta=0.210\pm0.005$.
%
\section{Conclusions}

To study the growth of the real contact area between rough self-affine manifolds 
we obtained statistically meaningful results for different cutoff wavelengths and Hurst exponents.
We demonstrated that the real contact area evolves nonlinearly with applied force even for reasonably 
small contact fractions $A/A_0 < 0.1$.
Note that an almost linear evolution was observed for rough contact between elasto-plastic materials \cite{persson2001prl,pei2005jmps,yastrebov2011cras}, 
however, in this case the Johnson's assumption~\cite{johnson1987b} on replacing two deformable solids by one solid with effective elasto-plastic
properties and superposed roughnesses is not verified and a full simulation of two deformable solids is required.

In conclusion, we state that to obtain realistic results in rough contact analysis, one needs to construct representative surfaces.
This representativity necessarily requires the long wavelength cutoff to be significantly smaller than the specimen size $\lambda_l \gtrapprox L/16$.
The Hurst exponent does not change the evolution of the real contact area with load if both the representativity and the smoothness of surfaces are maintained.
To describe the evolution of the real contact area we proposed a new phenomenological law which describes well our results for moderate pressures
and reduces to the classical BGT law~\cite{bush1975w} in the limit of infinitesimal contact fractions.
However, further work is required to verify the universality of this law.
In perspective we aim to study the pressure distribution and the evolution of the contact area up to the full contact,
for which the numerical results should be in a better agreement with Persson theory~\cite{persson2001jcp,persson2001prl}.
An important question, which we could not answer in our study, is how the bandwidth parameter influences the mechanical behavior of rough surfaces.
To address it, one needs to carry out numerical simulations on a considerably finer discretization of surfaces than was reported here. 
We hope that our results will motivate new simulations based on the notion of representativity and taking into account 
friction, visco-plasticity, adhesion and surface energy.

The financial support from the European Research Council (ERCstg UFO-240332) is greatly acknowledged.

\section{Appendix: representativity of rough surfaces}

The generation of realistic surfaces is the basis for a numerical analysis of the contact interaction between solids. It is widely acknowledged that real surfaces obey the property of self-affinity~\cite{johnson1987b,krim1995ijmpb} over many spatial scales.
Self-affine surfaces can be described by a power spectrum density (PSD) $C(k)$ following a power law of a wavenumber $k = L/\lambda$
\begin{equation}
 C(k) \sim k^{-2(H+1)}
\label{eq:HurstExponent}
\end{equation}
where $H$ is the so-called Hurst roughness exponent $H\in(0,1)$.
In reality the PSD may be considered confined between two cutoff wavenumbers:  $k_s$ and  $k_l$
corresponding to short and long wavelengths $\lambda_s$ and $\lambda_l$, respectively.
To be representative, a surface has to obey a statistically relevant distribution
of heights so that, its global mechanical response computed with periodic boundary conditions is equivalent to the 
response of any statistically similar rough surface. 
Since the mechanical response of the generated surfaces is affected by their height distributions and
the PSD, the surface generation procedure has to include both of them into consideration. 
While the typical PSD of self-affine surfaces is easily obtained using a random midpoint
algorithm \cite{fournier1982cacm} or by filtering a white noise~\cite{hu1992ijmtm}, 
the normality of the height distribution of simulated surfaces was, to the best of our knowledge, often ignored.

We will demonstrate that even if available numerical techniques permit to generate a self-affine surface
with a small error on the resulting PSD, the distribution of heights strongly depends 
on the range of wavelengths $[\lambda_s,\lambda_l]$ included in the spectrum. 
In order to quantify the deviation from a Gaussian distribution of heights we introduce the integral $L2$ error:
$$
  \varepsilon(S,N) = \sqrt{\int\limits_{-\infty}^{+\infty} \left|G(h,\bar h(S),\sigma(S)) - P_N^S(h)\right|^2\,dh}
$$
where the mean height $\bar h(S)$, the standard deviation $\sigma(S)$, and the probability density function (PDF) of heights $P_N^S$ are extracted from surface $S$ and
where $N$ is the number of bins used to evaluate the PDF of the surface. 
The function $G$ stands for the normal distribution 
$$
  G(h,\bar h, \sigma) = \frac{1}{\sqrt{2\pi} \sigma}\exp\left[\frac{-(h-\bar h)^2}{2 \sigma^2}\right]
$$
This measure of the surface representativity has been used in a parametric study to identify the
effect of the selected wavelength band $[\lambda_s,\lambda_l]$.

\begin{figure}[ht]
   \begin{center}
\includegraphics[width=0.8\textwidth]{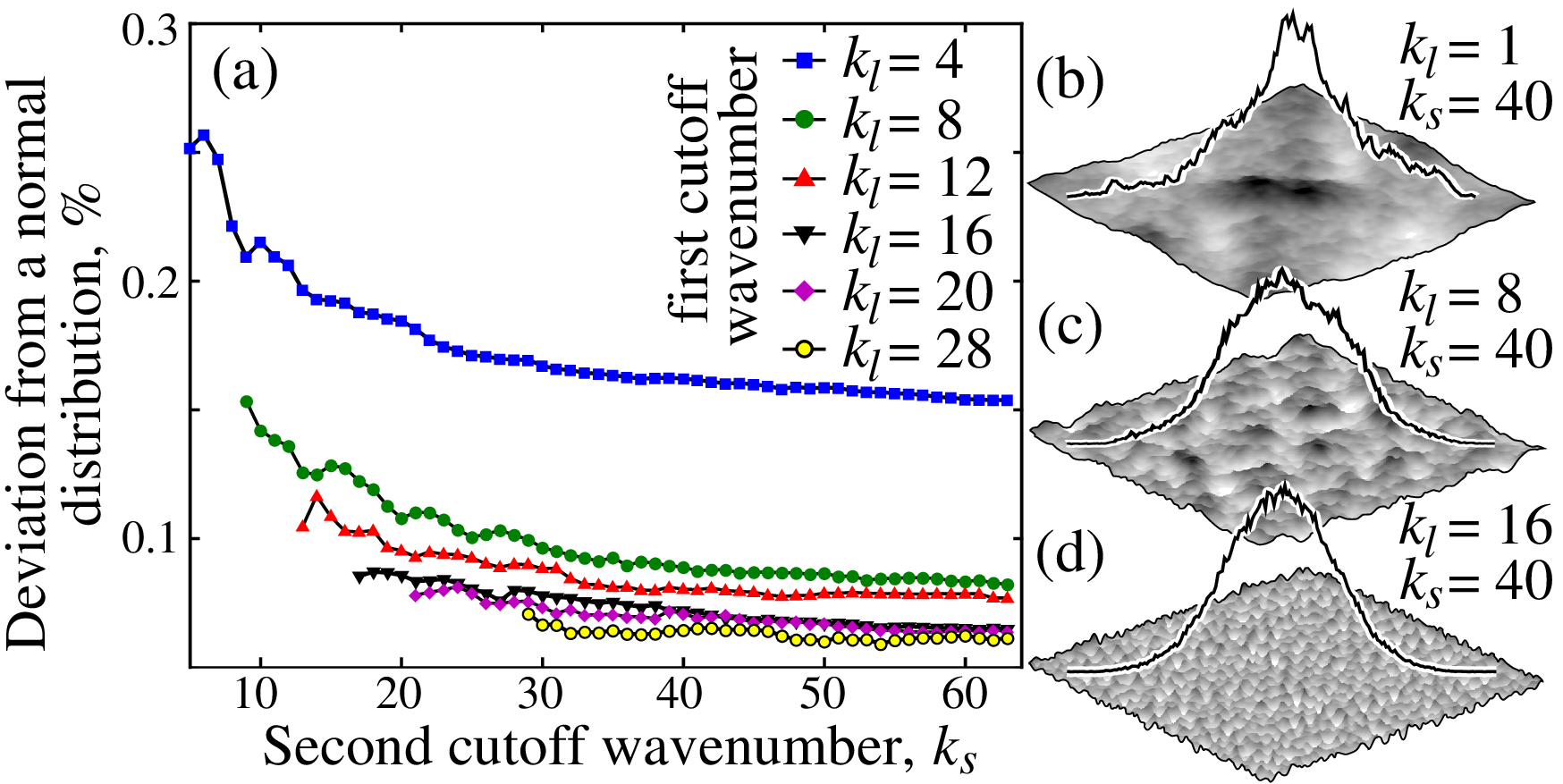}
\caption{\label{fig:app1}
Mean deviation from the Gaussian distribution (computed over 30 surfaces) with respect to the high cutoff wavenumber $k_s$. 
Various low cutoff wavenumber $k_l$ are presented by several curves. 
The Hurst roughness exponent for all surfaces is $H = 0.8$ and the root mean squared height is $\mbox{rms}=0.1$, each surface consists of $1024^2$ points.}
   \end{center}
\end{figure}

We have synthesized a large set of self-affine rough surfaces (30 per each combination of $\lambda_l$ and $\lambda_s$) containing $1024\times 1024$
points by using the filtering technique described in~\cite{hu1992ijmtm}. 
Fig.~\ref{fig:app1} represents the average error over generated surfaces for Hurst exponent $H=0.8$ and the number of bins $N=500$.
It can be observed that the error decreases with  increasing $k_s$ for all $k_l$.
However, the value of $k_l$ dictates the order of magnitude of the error for all $k_s$.
Therefore we may assume that for $k_l = 16$ and $k_s \ge 32$ the heights distributions of generated surfaces 
do not deviate significantly (in average) from the normal one. 

%
%

\section{Appendix: contact evolution law}

According to the simulations, the change of the mean contact pressure $\bar p = F/A$ with external pressure $p_0 = F/A_0$ decreases approximately as a power law of the real contact area fraction $A/A_0$:
\begin{equation}
 \label{eq:a1}
 \frac{\partial \bar p}{\partial p_0} = \beta\left(\frac{A_0}{A}\right)^\mu,\; \beta > 0,\;0 < \mu < 1
\end{equation}
In the limit of small contact we suppose the classical linear relation between $A/A_0$ and $p_0$ 
$$\frac{A}{A_0} = \frac{\kappa}{\sqrt{|\nabla h|^2} E^*} p_0.$$ 
However, in contrast to the Bush-Gibson-Thomas model~\cite{bush1975w} there is no need to analyze unreachably small fractions of the real contact area.
In the limit of the full contact the smoothness of the function is lost 
\begin{equation}
	\begin{cases}
		\frac{\partial\bar p}{\partial p_0} = \beta,&\mbox{ when }A \to A_0\mbox{ and }A_0-A > 0,\\
		\frac{\partial\bar p}{\partial p_0} = 1,&\mbox{ when }A = A_0
	\end{cases}
\end{equation}
In other words the slope of the real contact area is inversely proportional to the contact pressure $p_0^{\mbox{\tiny full}}$ needed to establish full contact with
the coefficient of proportionality  $d(A/A_0)/dp_0 = (1-\beta)/p_0^{\mbox{\tiny full}}$
however, this theory is supposed to be valid only in the range of moderate forces, far from the full contact range.
Solving Eq.~(\ref{eq:a1}) gives the following law for the evolution of the real contact area
$$
 \frac{A}{A_0} = \frac{1}{\left[ \beta + \left(
\frac{\sqrt{|\nabla h|^2} E^*}{\kappa p_0}\right)^{1-\mu} \right]^{1/(1-\mu)}},
$$
for the considered case we found $\kappa = (1.145\pm0.002)\sqrt{2\pi} \approx 2.870\pm0.005$, $\mu=0.55\pm0.02$, $\beta=0.210\pm0.005$.

\subsection{Details of the CEL derivation}

Starting with the observation that
\begin{equation}
  \label{eq:a2}
  \frac{\partial \bar p}{\partial p_0} = \beta\left(\frac{A_0}{A}\right)^\mu
\end{equation}
Developing the left part gives
$$
 \bar p = F/A, p_0 = F/A_0 \;\Rightarrow\; \frac{\partial \bar p}{\partial p_0} = A_0 \frac{A-F\partial A/\partial F}{A^2}
$$
Substituting this expression in the left part of (\ref{eq:a2}) gives
\begin{equation}
  \frac{A-F\frac{\partial A}{\partial F}}{A^2} = \frac{\beta}{A_0} \left(\frac{A_0}{A}\right)^\mu
\end{equation}
Grouping terms gives
$$
F\frac{\partial A}{\partial F} = A - \beta A_0\left(\frac{A_0}{A}\right)^{\mu-2} 
$$
Separating variables and integration of both sides reads as
\begin{equation}
\int\limits_{A_c}^A\frac{d (A/A_0)}{A/A_0 - \beta \left(A/A_0\right)^{2-\mu}} = \int\limits_{F_c}^{F} \frac{dF}{F}
\label{eq:a3} 
\end{equation}
$$
  \frac{1}{\mu-1}\ln\left(\frac{\left(A/A_0\right)^{\mu-1} - \beta}{ \left(A_c/A_0\right)^{\mu-1} -\beta}\right) = \ln(F/F_c)
$$  
Let $F_c \to 0,\;A_c \to 0$ and if at very small pressures the real contact area is proportional to the normal force with coefficient $\kappa$, then 
$$
  A_c = \frac{\kappa}{\sqrt{|\nabla h|^2} E^*} F_c
$$
Substituting this limit in the bottom boundary of the left integral in Eq.~(\ref{eq:a3}) and integrating, one obtains
$$
  \frac{1}{\mu-1}\ln\left(\frac{\left(A/A_0\right)^{\mu-1} - \beta}{ \left(\frac{\kappa}{\sqrt{|\nabla h|^2} E^*} F_c/A_0\right)^{\mu-1} -\beta}\right) = \ln(F/F_c)
$$
Since $\mu-1 < 0$ and $F_c$ tends to zero we can neglect $\beta$ in the denominator
$$
  \frac{1}{\mu-1}\ln\left(\frac{\left(A/A_0\right)^{\mu-1} - \beta}{ \left(\frac{\kappa}{\sqrt{|\nabla h|^2} E^*} F_c/A_0\right)^{\mu-1}}\right) = \ln(F/F_c)
$$
Taking the exponential of both parts one gets:
$$
 \frac{\left(A/A_0\right)^{\mu-1} - \beta}{ \left(\frac{\kappa}{\sqrt{|\nabla h|^2} E^*} F_c/A_0\right)^{\mu-1}} = (F/F_c)^{\mu-1}
$$
Now we can express the real contact area fraction $A/A_0$ as a function of the applied pressure $p_0 = F/A_0$:
$$
  \frac{A}{A_0} = \left[\beta +  \left(\frac{\kappa p_0}{\sqrt{|\nabla h|^2} E^*}\right)^{\mu-1}\right]^{1/(\mu-1)}
$$


\end{document}